\documentclass[onecolumn,prd,showpacs,preprintnumbers,floatfix£¬14pt]{revtex4}
\usepackage{dcolumn}
\usepackage{bm}
\usepackage{longtable}
\usepackage{graphicx,epsfig,latexsym,amssymb}
\usepackage{multirow,amsmath,array,booktabs,color}
\usepackage[section]{placeins}
\usepackage{makecell}

\begin{document}
%-------------------------------------------------------------------------------------------------------
\title{Photoproduction of the double $J/\psi$ ($\Upsilon$) at the LHC with forward proton tagging}

\author{Pan Xue-An$^1$}
\author{Li Gang$^1$}\email{ lig2008@mail.ustc.edu.cn}
\author{Song Mao$^1$}
\author{Zhang Yu$^{2,}$$^3$}
\author{Sun Hao$^4$}
\author{Guo Jian-You$^1$}

\affiliation{$^1$ School of Physics and Materials Science, Anhui University, Hefei, Anhui 230039, People's Republic of China}
\affiliation{$^2$ Institute of Physical Science and Information Technology, Anhui University, Hefei, Anhui 230039, People's Republic of China }
\affiliation{$^3$ CAS Center for Excellence in Particle Physics, Beijing 100049, China }
\affiliation{$^4$ Institute of Theoretical Physics, School of Physics, Dalian University of Technology, Dalian 116024, People's Republic of China}

\date{\today }
%-------------------------------------------------------------------------------------------------------
\begin{abstract}
 We calculate the photoproduction of double $J/\psi$ ($\Upsilon$) to leading order based on the nonrelativistic quantum chromodynamics factorization framework at the Large Hadron Collider with forward proton tagging. The numerical results of double $J/\psi$ photoproduction pp $\rightarrow$ p$\gamma$p $\rightarrow$ $J/\psi$ + $J/\psi$ with different forward detector acceptances ($\xi$) are presented.
  The total cross section of double $J/\psi$ photoproduction is less than 200 fb with 0.1 $<$ $\xi$ $<$ 0.5, but can reach about 1.37(1.27) pb with 0.0015 $<$ $\xi$ $<$ 0.5 ( 0.0015 $<$ $\xi$ $<$ 0.15 ). The double $J/\psi$ photoproduction may have the potential to be detected and provide an interesting signature; thus, it is useful for studying the mechanism of heavy quarkonium production. We also predict the double $\Upsilon$ photoproduction and find that they are, unfortunately, small (with less than 10 fb).
\end{abstract}

\maketitle

\section{Introduction}
The discovery of $J/\psi$ in 1974 opened up a new era of high energy physics: heavy quarkonium physics. In the early days, based on quantum chromodynamics (QCD), the color-singlet mechanism (CSM)\cite{C3} was used to explain the production and decay of heavy quarkonium. In CSM, the production of quarkonium is separated into the short-distance coefficients and a single nonperturbative parameter, in which the short-distance coefficients describe the Q\={Q} pair production, and the nonperturbative parameter describes the hadronization of the heavy Q\={Q} pair into the observable quarkonium state, which can be obtained by the bound state Bethe-Salpeter wave function. The Q\={Q} pair must be a color-singlet state with the same spin and angular momentum quantum numbers of the quarkonium state. Although the CSM has achieved a great deal of phenomenological success, there are still many theoretical predictions which are inconsistent with the experiments\cite{C7}, and the CSM also cannot handle the infrared divergences in P-wave and D-wave decay widths of heavy quarkonium\cite{C11}. The way to make up for these deficiencies of the CSM has been indicated by the work of Bodwin $\emph{et al.}$\cite{C2}, who have provided a theoretical framework, nonrelativistic quantum chromodynamics (NRQCD), describing the production and decay of the heavy quarkonium. Similarly to the CSM, the NRQCD also factorizes the production and decay of heavy quarkonium into two parts, the process-dependent short-distance coefficients (SDCs), which can be calculated based on the perturbation QCD, and the universal long-distance matrix elements (LDMEs) that can be extracted from experiments. Compared with the CSM, NRQCD allows the Q\={Q} pair to be a color-singlet or color-octet state. NRQCD has achieved great success in both theory and experiment, but there are still some challenges in interpreting the heavy quarkonium physics\cite{C6}. NRQCD is still under question and far from being proven; the phenomenological successes and challenges of NRQCD in describing the heavy quarkonium production and decay are nicely described in Refs.\cite{C8,C9,C10}. To deepen our understanding of the mechanism for heavy quarkonium production and decay, we should study more heavy quarkonium production and decay in the future.

The study of double heavy quarkonium production is one of the most interesting subjects in both theoretical\cite{D99,D14,D15,D19,D24,D2,D6,D51,D5,D4,D3,D1,D50,D7,D20,D21,D18,D8,D28,D22} and experimental\cite{D10,D16,D17,D100,D12,D13,D101,D23} physics, especially for the production of double $J/\psi$. The double $J/\psi$ production was first detected by the CERN-NA3 Collaboration in the 1980s\cite{D10}, and then tremendous progress has been made in systematic studies of double heavy quarkonium production. Many double heavy quarkonium productions, such as $J/\psi$ + $\chi_{c0}$\cite{D16} and $J/\psi$ + $\eta_c$\cite{D16,D17}, have been detected in $e^{+}e^{-}$ collision. However, the observed cross section for $J/\psi$ + $\eta_c$ production at B factories\cite{D17} is about an order of magnitude larger than the leading order (LO) NRQCD prediction; meanwhile, there is no evidence to show the production of double $J/\psi$ in $e^{+}e^{-}$ collisions at the Belle Collaboration\cite{D100}, which is conflict with the LO prediction\cite{D99,D14}. Finally, through efforts from the theoretical aspect, these phenomena can be explained by considering NLO QCD correction\cite{D15,D19} and the relativistic corrections\cite{D24}.  Recently, the hadroproduction of double $J/\psi$ has attracted considerable interest; measurements have been released by the LHCb\cite{D12} and CMS\cite{D13} collaborations at the LHC and by the D0 Collaborations\cite{D101} at the Tevatron. Many relevant theoretical works have been performed to predict\cite{D2} or interpret\cite{D6} these results. Now, the LO CSM\cite{D51} and NRQCD\cite{D5,D4} predictions and the NLO CSM\cite{D3} and partial NLO (NLO$^*$) contribution\cite{D1} are available. The LHCb measured cross section can be reasonably well described by the NLO result\cite{D3} in the single (hard) parton scattering (SPS) mechanism, but the CMS data exhibit very different behaviors with the NLO prediction. From the D0 and CMS measurements, though suffering from the uncertainty of the model dependence, Lansberg and Shao\cite{D50,D7} have found that the double parton scattering (DPS) contributions are dominant at large $\Delta$y and large $M_{\Psi\Psi}$, which can greatly reduce the discrepancy, but the ``CMS puzzle" is still an open question and challenges our understanding of the heavy quarkonium  production mechanism.

To reveal the mechanism of the heavy quarkonium production, in addition to the detailed studies at $e^+e^-$ collisions\cite{D20,D21,D18,D23} and hadron collisions\cite{D8,D28}, the interesting signature of double heavy quarkonium  has also been investigated through photoproduction at photon colliders\cite{D22}. The FP420 R$\&$D Collaboration put forward a plan\cite{E1} in 2009 to study standard model (SM) physics and to search for new physics signals. To achieve these goals, they need to reform the detectors in the LHC tunnel, so that these detectors can accurately measure very forward protons.  A significant fraction of pp collisions at the LHC will involve quasireal (low-$Q^{2}$) photon interactions, which are called the photon induced processes. The photon induced interactions have two processes at the LHC with the forward detector: (1) photon-photon processes and (2) photon-proton processes. In these processes, the photon is emitted by one (or both) incoming proton(s), and the photon will get energy $E_{\gamma}$ from the emitters proton. Therefore, the emitter proton should have some momentum fraction loss $\xi$, which is defined as the forward detector acceptance. The acceptance of $\xi$ is in the range 0.0015 $<$ $\xi_1$ $<$ 0.5 for the CMS-TOTEM forward detector, 0.1 $<$ $\xi_2$ $<$ 0.5 for the CMS-TOTEM forward detector, and 0.0015 $<$ $\xi_3$ $<$ 0.5 for the AFP-ATLAS forward detector. Compared with the usual pp or p\={p} collisions, $\gamma$$\gamma$ and $\gamma$p collisions can provide a cleaner environment. For this reason, we study the photoproduction of double $J/\psi$ ($\Upsilon$) at the LHC with three detector acceptances. General diagrams of double $J/\psi$ ($\Upsilon$) production by the photon induced processes at the LHC are presented in Fig. 1: (1) pp $\rightarrow$ p$\gamma$$\gamma$p $\rightarrow$ $\mathcal{Q}\mathcal{Q}$ (left panel) corresponds to the photon-photon ($\gamma$$\gamma$) interaction where photons radiate from both protons and produce double heavy quarkonia $\mathcal{Q}$ = $J/\psi$ or $\Upsilon$. (2) pp $\rightarrow$ p$\gamma$p $\rightarrow$ $\mathcal{Q}\mathcal{Q}$ (right panel) refers to photoproduction or photon-proton ($\gamma$p) production: a photon emitted from the proton induces deep inelastic scattering with the incoming proton and produce double heavy quarkonia.  Although the $\gamma\gamma$ process is cleaner than the $\gamma$p process, the $\gamma$p process has higher energy and effective luminosity, so the study of the $\gamma$p process is more attractive than the $\gamma\gamma$ process, and the $\gamma$p process has been studied in many phenomenological studies\cite{B8,E2,E3,E4,E5,E6}. In this paper, we calculate the LO cross section of double $J/\psi$ ($\Upsilon$) photoproduction for the pp $\rightarrow$ p$\gamma$p $\rightarrow$ $\mathcal{Q}$ + $\mathcal{Q}$ process at the LHC with forward detector acceptances.

The paper is organized as follows. In Sec.~II, we introduce the calculation framework for the LO cross section of double $J/\psi$ ($\Upsilon$) at the LHC with forward detector acceptances in detail. In Sec.~III, we present the input parameters and numerical results. A short summary and discussions are given in Sec. IV.

\begin{figure}[!htbp]
\vspace*{-3cm}
\hspace*{-6cm}
   \includegraphics[scale=1]{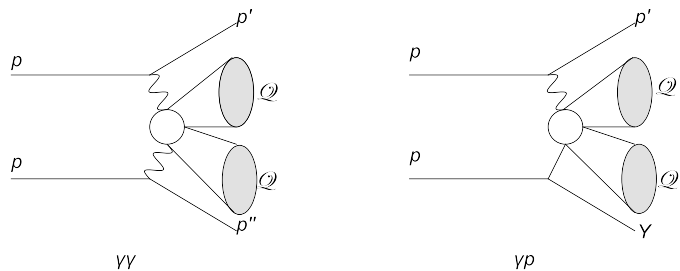}\newline
   \caption{General diagrams of heavy quarkonium pair production for the photon induced production at the CERN LHC: pp $\rightarrow$ p$\gamma\gamma$p $\rightarrow$ $\mathcal{Q}\mathcal{Q}$ (left) and pp $\rightarrow$ p$\gamma$p $\rightarrow$ $\mathcal{Q}\mathcal{Q}$ (right).  }
   \end{figure}

\section{the details of the calculation framework}
In this section, we introduce the calculation framework for the process pp $\rightarrow$ p$\gamma$p $\rightarrow$ $\mathcal{Q}$ + $\mathcal{Q}$ based on the NRQCD factorization formalism to LO at the LHC with forward detector acceptances in detail, where $\mathcal{Q}$ is $J/\psi$ or $\Upsilon$. The cross section for the pp $\rightarrow$ p$\gamma$p $\rightarrow$ $\mathcal{Q}$ + $\mathcal{Q}$ process can be expressed as
\begin{eqnarray}
\sigma( pp \rightarrow p\gamma p \rightarrow \mathcal{Q} + \mathcal{Q} )  =
  \int d{\xi}d{x}[ f_{\gamma/A}(\xi)G_{g/B}(x,\mu_{f})\sum\limits_{n_{1},n_{2}}
\hat{\sigma}(\gamma g \rightarrow Q \overline{Q}[n_{1}] + Q \overline{Q}[n_{2}] ) \notag \\
\times \left\langle {O_{1,8}^{\mathcal{Q}}[n_{1}]}\right\rangle \left\langle {O_{1,8}^{\mathcal{Q}}[n_{2}]}\right\rangle + A \leftrightarrow B ]
\end{eqnarray}
where $\left\langle {O_{1,8}^{\mathcal{Q}}[n]}\right\rangle$ is the long-distance matrix element, which describes the hadronization of the heavy Q\={Q}[$n$] pair into the observable quarkonium state $\mathcal{Q}$, and the Fock states $n_{1}$, $n_{2}$ =$\sideset{^3}{^{[1]}_1}{\mathop{{S}}}$, $\sideset{^1}{^{[8]}_0}{\mathop{{S}}}$, $\sideset{^3}{^{[8]}_1}{\mathop{{S}}}$, $\sideset{^3}{^{[8]}_J}{\mathop{{P}}}$. The $\hat{\sigma}(\gamma g \rightarrow Q \overline{Q}[n_{1}] + Q \overline{Q}[n_{2}] )$ denotes the short-distance cross section for the partonic process $\gamma g \rightarrow Q \overline{Q}[n_{1}] + Q \overline{Q}[n_{2}]$, which is obtained by applying the covariant projection method. $G_{g/B}(x)$ denotes the gluon parton density function, and $x$ is the momentum fraction of the proton momentum carried by the gluon. $\xi$ = $E_\gamma/E$ is the ratio between scattered low-$Q^2$ photons $E_\gamma$ and incoming proton energy $E$. In our calculation, $f_{\gamma/A}(\xi)$ is the effective photon density function which is described by the equivalent photon approximation (EPA)\cite{E3,B1},
\begin{eqnarray}
f_{\gamma/A}(\xi) =\int\limits_{Q_{\min }^2}^{Q_{\max }^2} \frac{{d{N_\gamma(\xi) }}}{{d{\xi}d{Q^2}}}d {Q^2},
\end{eqnarray}
and $d{N_\gamma}(\xi)$ is the spectrum of quasireal photons\cite{E3,B1} :
\begin{eqnarray}
\frac{{d{N_\gamma}}}{{d{E_\gamma}d{Q^2}}}& = & \frac{\alpha}{\pi}\frac{1}{E_{\gamma}Q^{2}}[(1-\frac{E_{\gamma}}{E})(1-%
\frac{Q_{min}^{2}}{Q^{2}})F_{E}+\frac{E_{\gamma}^{2}}{2E^{2}}F_{M}].
\end{eqnarray}
With $\xi$ = $E_\gamma/E$, we can obtain the expression as follows:
\begin{eqnarray}
\frac{{d{N_\gamma(\xi) }}}{{d{\xi}d{Q^2}}} &=& \frac{\alpha}{\pi}\frac{1}{\xi Q^{2}}[(1 - \xi)(1-%
\frac{Q_{min}^{2}}{Q^{2}})F_{E}+\frac{\xi^{2}}{2}F_{M}],
\end{eqnarray}
where
\begin{eqnarray}
Q_{\min }^2& = & \frac{{m_{p}^{2}E_{\gamma}^{2}}}{E(E - {E_{\gamma})}} = \frac{{m_{p}^{2}\xi^{2}}}{1 - \xi} ,
F_{E}  =  \frac{4m_{p}^{2}G_{E}^{2}+Q^{2}G_{M}^{2}}{4m_{p}^{2} + Q^{2}}, \notag \\
G_{E}^{2} & = & \frac{G_{M}^{2}}{\mu_{p}^{2}}=(1+\frac{Q^{2}}{Q_{0}^{2}})^{-4},
 F_{M}  =  G_{M}^{2},
\end{eqnarray}
$\alpha$ is the fine-structure constant, $m_{p}$ is the mass of the proton, $Q_{0}^{2}$  =  0.71 GeV$^{2}$, $\mu_{p}^{2}$ = 7.78 is the magnetic moment of the proton, and the range of $Q_{\max}^{2}$ is 2 $\sim$ 4 GeV$^{2}$. The $f_{\gamma/A}(\xi)$ integral expression in Eq.(2) can also be written as\cite{E3,B1}
\begin{eqnarray}
f_{\gamma/A}(\xi) = \frac{\alpha }{\pi}\frac{1}{\xi}(1 - \xi)[\varphi (\frac{{Q_{\max }^2}}{{Q_0^2}}) - \varphi (\frac{{Q_{\min }^2}}{{Q_0^2}})],
\end{eqnarray}
where the function $\varphi$ is defined as
\begin{eqnarray}
\varphi (\nu) & = & (1 + ay)[ - \rm{ln}(1 + {\nu^{ - 1}}) + \sum\limits_{k = 1}^3
{\frac{1}{{k{{(1 + \nu)}^k}}}}] \notag \\
&& +\frac{{(1 - b)y}}{{4\nu{{(1 + \nu)}^3}}} + c(1 + \frac{y}{4}) {\rm{ [ln}}\frac{{1 + \nu - b}}{{1 + \nu}} + \sum\limits_{k = 1}^3 {\frac{{{b^k}}}{{k{{(1 + \nu)}^k}}}]},
\end{eqnarray}
with
\begin{eqnarray}
y & = &\frac{{\xi^2}}{{1 - \xi}} , a = \frac{1}{4}(1 + \mu _p^2) + \frac{{4m_p^2}}{{Q_0^2}} \notag \\
b & = & 1 - \frac{{4m_p^2}}{{Q_0^2}}, c  =  \frac{{\mu _p^2 - 1}}{{{b^4}}}.
\end{eqnarray}

Finally, the total cross section for the pp $\rightarrow$ p$\gamma$p $\rightarrow$ $\mathcal{Q}$ + $\mathcal{Q}$ process can be expressed as
\begin{eqnarray}
\sigma(pp \rightarrow p\gamma p \rightarrow \mathcal{Q} + \mathcal{Q}) =
\int\limits_{\frac{{4m}}{{\sqrt s }}}^{\sqrt {{\xi _{\max }}}}{2zdz\int\limits_{Max({z^2},{\xi_{\min }})}^{{\xi _{\max }}} {\frac{{d{\xi}}}{{{\xi}}}f_{\gamma/A}({\xi})} {G_{g/B}}} (\frac{z}{{{\xi}}},
{\mu _f})\sum\limits_{n_{1},n_{2}}
\hat{\sigma}(\gamma g \rightarrow Q \overline{Q}[n_{1}] + Q \overline{Q}[n_{2}]) \notag \\
\times\left\langle {\mathcal{O}_{1,8}^{\mathcal{Q}}[n_{1}]}\right\rangle \left\langle {\mathcal{O}_{1,8}^{\mathcal{Q}}[n_{2}]}\right\rangle
+ A \leftrightarrow B,
\end{eqnarray}
where $z^2$ = $\xi$$x$, $\xi _{\min}$ ($\xi _{\max}$) is the lower (upper) limit of forward detector acceptance; $s$ is the square of center-of-mass energy of the collider, and $m$ is the mass of the quark.

The LO short-distance cross section of the partonic process $\gamma(p_1) + g(p_2) \rightarrow Q \overline{Q}[n_{1}](p_3) + Q\overline{Q}[n_{2}](p_4)$ is presented as follows:
\begin{eqnarray}
\hat{\sigma}(\gamma g \rightarrow Q \overline{Q}[n_{1}] + Q \overline{Q}[n_{2}])
 = \frac{1}{16\pi\hat{s}^{2}N_{1col}N_{1pol}N_{2col}N_{2pol}}
 \int \limits_{\hat{t}_{min}}^{\hat{t}_{max}}d\hat{t} \overline{\sum}\left|\mathcal{A}_{S,L}\right|^2.
\end{eqnarray}
The Mandelstam variables are defined as $\hat{s} = (p_1 + p_2)^2$, $\hat{t} = (p_1 - p_3)^2$, $\hat{u} = (p_1 - p_4)^2$. $N_{1col}$ ($N_{2col}$) refers to the number of color states of the $Q\overline{Q}[n_{1}]$ ($Q\overline{Q}[n_{2}]$). $N_{1pol}$ ($N_{2pol}$) denotes the number of polarization states of the $Q\overline{Q}[n_{1}]$ ($Q\overline{Q}[n_{2}]$). The summation is taken over the spins and colors of the final states; the bar denotes averaging over the spins and colors of initial partons, and being divided by the identical particle factor. The $\mathcal{A}_{S,L}$ can be obtained from Ref.\cite{B7}:
\begin{eqnarray}
\mathcal{A}_{S=0,L=0} & = & Tr[\mathcal{C}\Pi_{0}\mathcal{A}]|_{q=0}, \notag \\
\mathcal{A}_{S=1,L=0} & = &\epsilon_{\alpha} Tr[\mathcal{C}\Pi_{1}^{\alpha}\mathcal{A}]|_{q=0}, \nonumber \\
\mathcal{A}_{S=0,L=1} & = &\epsilon_{\alpha}\frac{d}{dq_{\alpha}} Tr[\mathcal{C}\Pi_{0}\mathcal{A}]|_{q=0}, \notag \\
\mathcal{A}_{S=1,L=1} & = &\epsilon_{\alpha\beta} \frac{d}{dq_{\beta}}Tr[\mathcal{C}\Pi_{1}^{\alpha}\mathcal{A}]|_{q=0}.
\end{eqnarray}
where $\mathcal{A}$ refers to the QCD amplitude with amputated heavy-quark spinors and the lower $q$ denotes the momentum of the heavy-quark in the $Q\overline{Q}$ rest frame.
$\Pi_{0/1}$ are the spin projectors onto the spin-singlet and spin-triplet states as
\begin{eqnarray}
\Pi_{0} & = & \frac{1}{\sqrt{8m^3}}(\frac{ /\kern-0.57em P}{2} - /\kern-0.50em q - m)\gamma_{5}(\frac{ /\kern-0.57em P}{2} + /\kern-0.50em q + m), \notag \\
\Pi_{1} & = & \frac{1}{\sqrt{8m^3}}(\frac{ /\kern-0.57em P}{2} - /\kern-0.50em q - m)\gamma^{\alpha}(\frac{ /\kern-0.57em P}{2} + /\kern-0.50em q + m),
\end{eqnarray}
where the momentum $\emph{P}$ is the total momentum of the heavy quarkonium.

$\mathcal{C}$ is the color projection operators with the following expression:
\begin{eqnarray}
\mathcal{C}_{1} &=& \frac{\delta_{ij}}{\sqrt{N_{c}}} \notag \\
\mathcal{C}_{8} &=& \sqrt{2}T_{ij}^{c}
\end{eqnarray}
where $\mathcal{C}_{1}, \mathcal{C}_{8}$ represent the color-singlet and color-octet states, respectively. $\epsilon_{\alpha}$ and $\epsilon_{\alpha\beta}$ denote the polarization vector and tensor of the $Q\overline{Q}$ states, respectively. The summation over the polarization is
\begin{eqnarray}
\sum \limits_{J_z}\epsilon_{\alpha}\epsilon_{\alpha'}^{*} & = & \Pi_{\alpha\alpha'}, \notag \\
\sum \limits_{J_z}\epsilon_{\alpha\beta}^{0}\epsilon_{\alpha'\beta'}^{0*} & = & \frac{1}{3}\Pi_{\alpha\beta}\Pi_{\alpha'\beta'}, \notag \\
\sum \limits_{J_z}\epsilon_{\alpha\beta}^{1}\epsilon_{\alpha'\beta'}^{1*} & = & \frac{1}{2}(\Pi_{\alpha\alpha'}\Pi_{\beta\beta'}-\Pi_{\alpha\beta'}\Pi_{\alpha'\beta}), \notag \\
\sum \limits_{J_z}\epsilon_{\alpha\beta}^{2}\epsilon_{\alpha'\beta'}^{2*} & = & \frac{1}{2}(\Pi_{\alpha\alpha'}\Pi_{\beta\beta'}+\Pi_{\alpha\beta'}\Pi_{\alpha'\beta})
-\frac{1}{3}\Pi_{\alpha\beta}\Pi_{\alpha'\beta'}
\end{eqnarray}
with
\begin{eqnarray}
\Pi_{\alpha\beta} = -g_{\alpha\beta}+\frac{P_{\alpha}P_{\beta}}{M^2},
\end{eqnarray}
and $M$ is the mass of the heavy quarkonium.

We use FeynArts\cite{B6} to generate Feynman diagrams and amplitudes for the partonic process $\gamma(p_1)$ + $g(p_2)$ $\rightarrow Q\overline{Q}[n_1](p_3)$ + $Q\overline{Q}[n_2](p_4)$, and we further reduce the Feynman amplitudes using FeynCalc\cite{B3} and FeynCalcFormLink\cite{B4}. There are 48 Feynman diagrams for this partonic process at tree level; some Feynman diagrams are shown in Fig. 2.
\begin{figure}[!htbp]
\centering
\vspace*{-4cm}
\hspace*{-2cm}
 \includegraphics[scale=0.85]{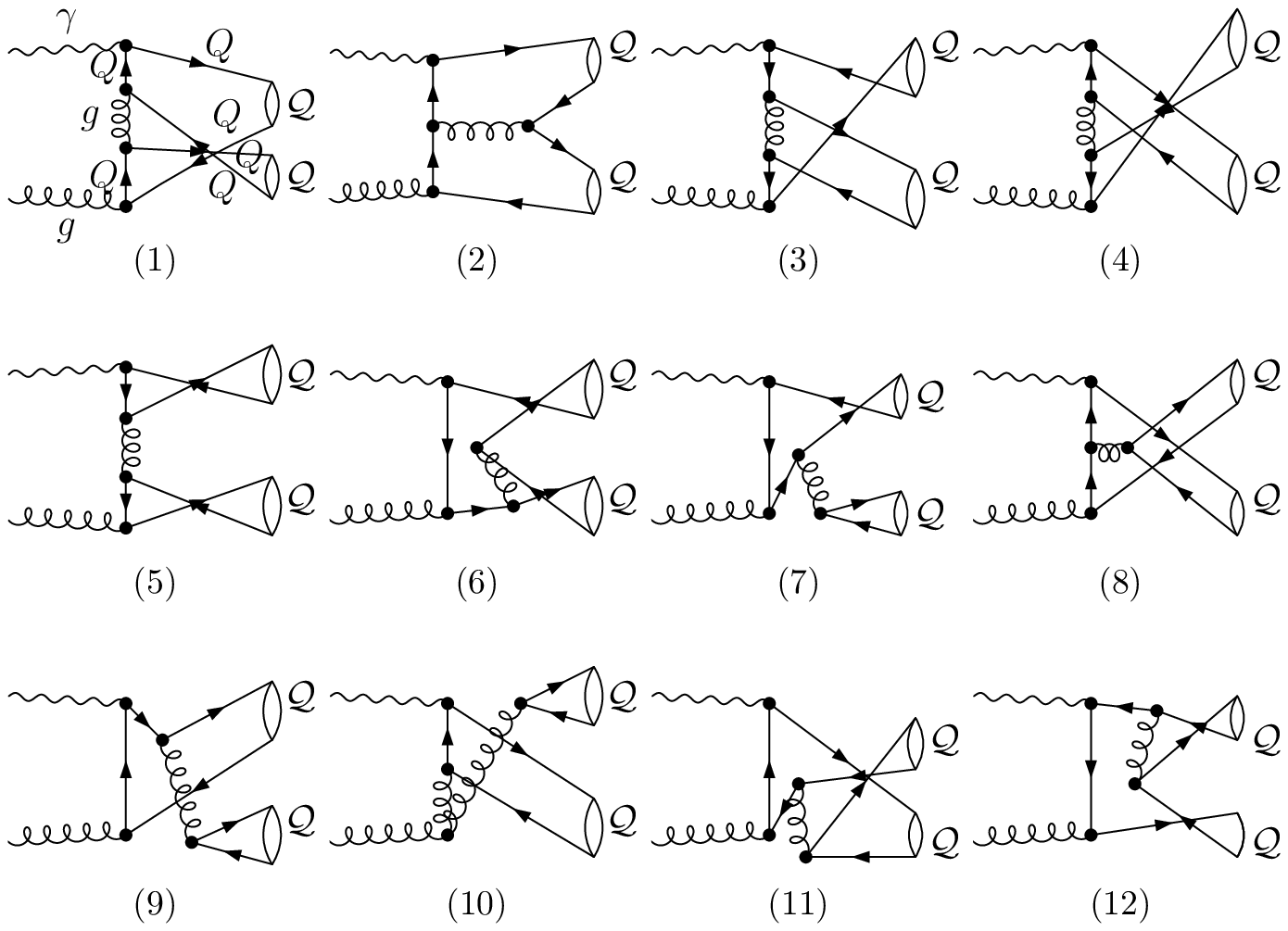}\newline
   \vspace*{-10cm}\caption{ Some Feynman diagrams at LO for the partonic process $\gamma(p_1)$ + $g(p_2)$ $\rightarrow Q\overline{Q}[n_1](p_3)$ + $Q\overline{Q}[n_2](p_4)$.[The sine, helical, and straight lines represent photons, gluons, and (anti)quarks respectively.]  }
\end{figure}

\section{numerical results}
In the numerical calculation, we take $m_{p}$ = 0.938272046 GeV as the mass of the proton, $Q_{\max}^{2}$ = 4 GeV$^2$. We set the constraints $p_{T}$ $>$ 3 GeV for $\mathcal{Q}$; the masses of the heavy quark are set as $m_{c}$ = 1.5 GeV, $m_{b}$ = 4.75 GeV; and the mass of $J/\psi$ and $\Upsilon$ are 2$m_{c}$ and 2$m_{b}$, respectively. The colliding energy is $\sqrt{s}$ = 14 TeV. The factorization scale is chosen as $\mu_f$ = $\sqrt{p_{T}^{2}+M^2}$, where $p_{T}$ is the transverse momentum of $\mathcal{Q}$. We use the CTEQ6L1\cite{A3} parton distribution function as the default and the $\alpha_{s}$ is extracted from the PDFs. Numerical calculations are performed by FormCalc\cite{B5}.

In this paper, we focus on photoproduction of the double $J/\psi$ ($\Upsilon$) at the LHC with forward detector acceptances. We choose the LDMEs for $J/\psi$ from Ref.\cite{A1} as

\begin{eqnarray}
\left\langle {\mathcal{O}^{J/\psi}[\sideset{^3}{^{[1]}_1}{\mathop{{S}}}]}\right\rangle &=& 1.1 \;\mathrm{GeV^3}, \notag \\
\left\langle {\mathcal{O}^{J/\psi}[\sideset{^1}{^{[8]}_0}{\mathop{{S}}}]}\right\rangle &=& 1\times 10^{-2} \;\mathrm{GeV^3}, \notag \\
\left\langle {\mathcal{O}^{J/\psi}[\sideset{^3}{^{[8]}_1}{\mathop{{S}}}]}\right\rangle &=& 1.12\times 10^{-2} \;\mathrm{GeV^3}, \notag \\
\left\langle {\mathcal{O}^{J/\psi}[\sideset{^3}{^{[8]}_0}{\mathop{{P}}}]}\right\rangle &=& 11.25\times 10^{-3} \;\mathrm{GeV^5},
\end{eqnarray}
and the LDMEs for $\Upsilon$ from Ref.\cite{A2} as
\begin{eqnarray}
\left\langle {\mathcal{O}^{\Upsilon}[\sideset{^3}{^{[1]}_1}{\mathop{{S}}}]}\right\rangle &=& 9.28 \;\mathrm{GeV^3}, \notag \\
\left\langle {\mathcal{O}^{\Upsilon}[\sideset{^1}{^{[8]}_0}{\mathop{{S}}}]}\right\rangle &=& 2\times 10^{-2} \;\mathrm{GeV^3}, \notag \\
\left\langle {\mathcal{O}^{\Upsilon}[\sideset{^3}{^{[8]}_1}{\mathop{{S}}}]}\right\rangle &=& 15\times 10^{-2} \;\mathrm{GeV^3}, \notag \\
\left\langle {\mathcal{O}^{\Upsilon}[\sideset{^3}{^{[8]}_0}{\mathop{{P}}}]}\right\rangle &=& 45.125\times 10^{-2} \;\mathrm{GeV^5}.
\end{eqnarray}
 For $\left\langle {\mathcal{O}^{\mathcal{Q}}[\sideset{^3}{^{[8]}_J}{\mathop{{P}}}]}\right\rangle$ with $J$ = 0, 1, 2, due to the spin symmetry of heavy quark, we have the relations
 \begin{eqnarray}
 \left\langle {\mathcal{O}^{\mathcal{Q}}[\sideset{^3}{^{[8]}_J}{\mathop{{P}}}]}\right\rangle &=& (2J+1)\left\langle {\mathcal{O}^{\mathcal{Q}}[\sideset{^3}{^{[8]}_0}{\mathop{{P}}}]}\right\rangle
 \end{eqnarray}

 In the case of the Fock states, $n_1$ and $n_2$ are both color-singlet states, the LO photoproduction of double $J/\psi$ is forbidden due to the color conservation, and only the Fock states $n_1$ and $n_2$ with at least one color-octet state may contribute to the double $J/\psi$ photoproduction. In our calculation, the contributions from the $\sideset{^3}{^{[1]}_1}{\mathop{{S}}}$ + $\sideset{^3}{^{[8]}_J}{\mathop{{P}}}$ channel to the double $J/\psi$ photoproduction vanish due to the C-parity conservation. As there are only two colored particles (gluon and $\sideset{^3}{^{[8]}_J}{\mathop{{P}}}$), color can be effectively dropped out. Then the gluon can effectively have the same C-parity as the photon, and the $\sideset{^3}{^{[8]}_J}{\mathop{{P}}}$ can effectively have the same C-parity as the $\sideset{^3}{^{[1]}_J}{\mathop{{P}}}$. For the same reason, the $\sideset{^3}{^{[1]}_1}{\mathop{{S}}}$ + $\sideset{^1}{^{[8]}_0}{\mathop{{S}}}$ channel also contributes nothing to the double $J/\psi$ photoproduction. In Table I, we present our theoretical prediction for the LO cross section of double $J/\psi$ at the LHC with forward detector acceptances. From the table, we can see that at the CMS-TOTEM forward detector with $\xi_2$ the cross section of double $J/\psi$ is much less than other two forward detector acceptances $\xi_1$ and $\xi_3$  because the effective photon density function $f_{\gamma}(\xi)$ is heavily suppressed when $\xi$ $>$ 0.1. As for the $\xi$ = $\xi_1$ and $\xi$ = $\xi_3$, the cross section of double $J/\psi$ can reach the magnitude of pb with the effective photon density function $f_{\gamma}(\xi)$ enhanced in the small $\xi$ region; the cross section of double $J/\psi$ is about 1369.06 and 1268.32 fb for detector acceptances with $\xi_1$ and $\xi_3$, respectively. From our results, we can see that the signal of double $J/\psi$ photoproduction at the LHC with forward detector acceptances mostly comes from the small $\xi$ region. To illustrate the dependence of the cross section on the parameter $\xi$, we present the distribution of $d\sigma/d log(\xi)$ for the double $J/\psi$ photoproduction at the LHC with the forward detector acceptances in Fig.$\;$3.

\begin{table}[!htbp]
\caption{The LO cross section (fb) for the double $J/\psi$ production at the LHC with forward detector acceptances.}
\begin{tabular}{|c|c|c|c|}
\hline
        & 0.0015 $<$ $\xi_1$ $<$ 0.5 & 0.1 $<$ $\xi_2$ $<$ 0.5   & 0.0015 $<$ $\xi_3$ $<$ 0.15 \\
\hline
$\sigma(pp \rightarrow p\gamma p \rightarrow J/\psi + J/\psi )$& 1369.06& 165.86& 1268.32\\
\hline
\end{tabular}
\end{table}

\begin{figure}[!htbp]
\centering
\hspace*{2cm}
   \includegraphics[scale=0.26]{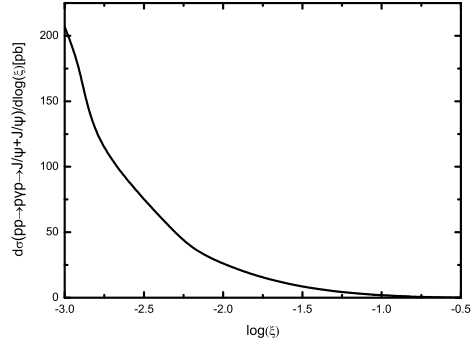}\newline
  \caption{ The distribution of $d\sigma/d log(\xi)$ for the double $J/\psi$ photoproduction at the LHC with forward detector acceptances}
 \end{figure}

In Fig.$\;$4, we present the LO distributions of $p_{T}^{J/\psi}$ (left panel) and $y^{J/\psi}$ (right panel) for the process pp $\rightarrow$ p$\gamma$p $\rightarrow$ $J/\psi$ + $J/\psi$ at the CMS-TOTEM forward detector with $\xi_1$. From this figure, we can see that when $p_{T}^{J/\psi}$ is less than about 12 GeV, the $\sideset{^3}{^{[1]}_1}{\mathop{{S}}}$ + $\sideset{^3}{^{[8]}_1}{\mathop{{S}}}$ and $\sideset{^3}{^{[8]}_1}{\mathop{{S}}}$ + $\sideset{^3}{^{[1]}_1}{\mathop{{S}}}$ channels give the main contribution to the $p_T$ distribution of photoproduction processes; with the $p_{T}^{J/\psi}$ increasing, the contributions of the $\sideset{^3}{^{[1]}_1}{\mathop{{S}}}$ + $\sideset{^3}{^{[8]}_1}{\mathop{{S}}}$ and $\sideset{^3}{^{[8]}_1}{\mathop{{S}}}$ + $\sideset{^3}{^{[1]}_1}{\mathop{{S}}}$ channels decrease rapidly, and the differential cross section is dominated by the $\sideset{^3}{^{[8]}_1}{\mathop{{S}}}$ + $\sideset{^3}{^{[8]}_J}{\mathop{{P}}}$ and $\sideset{^3}{^{[8]}_J}{\mathop{{P}}}$ + $\sideset{^3}{^{[8]}_1}{\mathop{{S}}}$ channel contributions in the large $p_T$ area. In the range of 3 GeV $<$ $p_{T}^{J/\psi}$ $<$ 10 GeV, the d$\sigma/dp_{T}^{J/\psi}$ is in the range of [2.75, 1255.73] fb/GeV, and it reaches the highest point at $p_{T}^{J/\psi}$ $\approx$ 3 GeV; then the d$\sigma/dp_{T}^{J/\psi}$ decrease fast with the $p_{T}^{J/\psi}$ increasing.

\begin{figure}[!htbp]
   \includegraphics[scale=0.26]{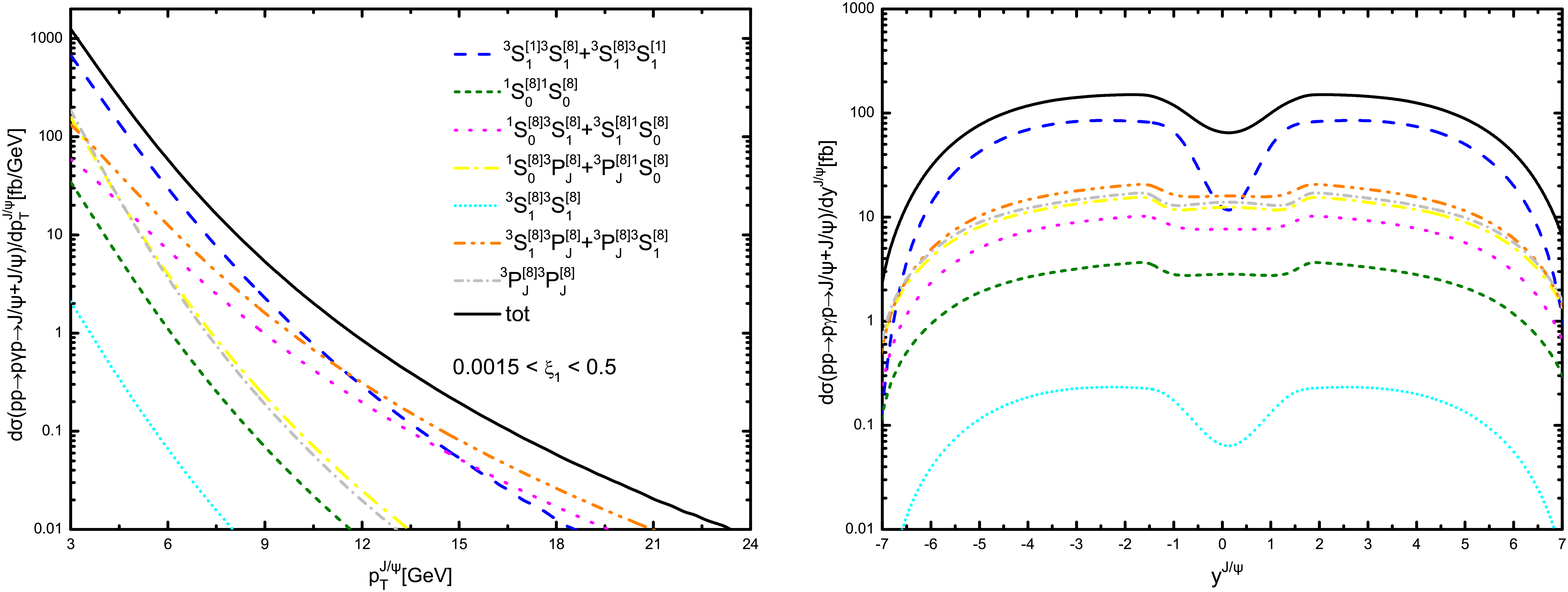}\newline
   \hspace*{5cm}
   \caption{ The LO distributions of $p_{T}^{J/\psi}$ (left) and $y^{J/\psi}$ (right) for the process pp $\rightarrow$ p$\gamma$p $\rightarrow$ $J/\psi$ + $J/\psi$ and the contributions of the $\sideset{^3}{^{[1]}_1}{\mathop{{S}}}$$\sideset{^3}{^{[8]}_1}{\mathop{{S}}}$ + $\sideset{^3}{^{[8]}_1}{\mathop{{S}}}$$\sideset{^3}{^{[1]}_1}{\mathop{{S}}}$ (blue dashed curve), $\sideset{^1}{^{[8]}_0}{\mathop{{S}}}$$\sideset{^1}{^{[8]}_0}{\mathop{{S}}}$ (olive short dashed curve), $\sideset{^1}{^{[8]}_0}{\mathop{{S}}}$$\sideset{^3}{^{[8]}_1}{\mathop{{S}}}$ + $\sideset{^3}{^{[8]}_1}{\mathop{{S}}}$$\sideset{^1}{^{[8]}_0}{\mathop{{S}}}$(magenta dotted curve), $\sideset{^1}{^{[8]}_0}{\mathop{{S}}}$$\sideset{^3}{^{[8]}_J}{\mathop{{P}}}$ + $\sideset{^3}{^{[8]}_J}{\mathop{{P}}}$$\sideset{^1}{^{[8]}_0}{\mathop{{S}}}$(yellow dashed-dotted curve), $\sideset{^3}{^{[8]}_1}{\mathop{{S}}}$$\sideset{^3}{^{[8]}_1}{\mathop{{S}}}$ (cyan short dotted curve), $\sideset{^3}{^{[8]}_1}{\mathop{{S}}}$$\sideset{^3}{^{[8]}_J}{\mathop{{P}}}$ + $\sideset{^3}{^{[8]}_J}{\mathop{{P}}}$$\sideset{^3}{^{[8]}_1}{\mathop{{S}}}$ (orange dashed-dotted-dotted curve), $\sideset{^3}{^{[8]}_J}{\mathop{{P}}}$$\sideset{^3}{^{[8]}_J}{\mathop{{P}}}$ (LT gray short dashed-dotted curve) channels, and total (black solid curve) at the CMS-TOTEM forward detector with 0.0015 $<$ $\xi_1$ $<$ 0.5.    }
   \end{figure}

The figures for the LO distributions of $p_{T}^{J/\psi}$ (left panel) and $y^{J/\psi}$ (right panel) for the process pp $\rightarrow$ p$\gamma$p $\rightarrow$ $J/\psi$ + $J/\psi$ at the CMS-TOTEM forward detector with $\xi_2$ are drawn in Fig. 5. From the figure, we can see that the shape of the $p_{T}^{J/\psi}$ distribution looks similar to that in Fig. 4; the $\sideset{^3}{^{[1]}_1}{\mathop{{S}}}$ + $\sideset{^3}{^{[8]}_1}{\mathop{{S}}}$ and $\sideset{^3}{^{[8]}_1}{\mathop{{S}}}$ + $\sideset{^3}{^{[1]}_1}{\mathop{{S}}}$ channels also give the main contribution for the double $J/\psi$ photoproduction process in the small $p_{T}^{J/\psi}$ region, and the $\sideset{^3}{^{[8]}_1}{\mathop{{S}}}$ + $\sideset{^3}{^{[8]}_J}{\mathop{{P}}}$ and $\sideset{^3}{^{[8]}_J}{\mathop{{P}}}$ + $\sideset{^3}{^{[8]}_1}{\mathop{{S}}}$ channel contributions dominate production in the large $p_{T}^{J/\psi}$ region. The d$\sigma/dp_{T}^{J/\psi}$ reaches the maximum value 147.44 fb/GeV with $p_{T}^{J/\psi}$ $\approx$ 3 GeV; then d$\sigma/dp_{T}^{J/\psi}$ quickly decreases to less than 1 fb/GeV at $p_{T}^{J/\psi}$ $\approx$ 10 GeV.

\begin{figure}[!htbp]
   \includegraphics[scale=0.26]{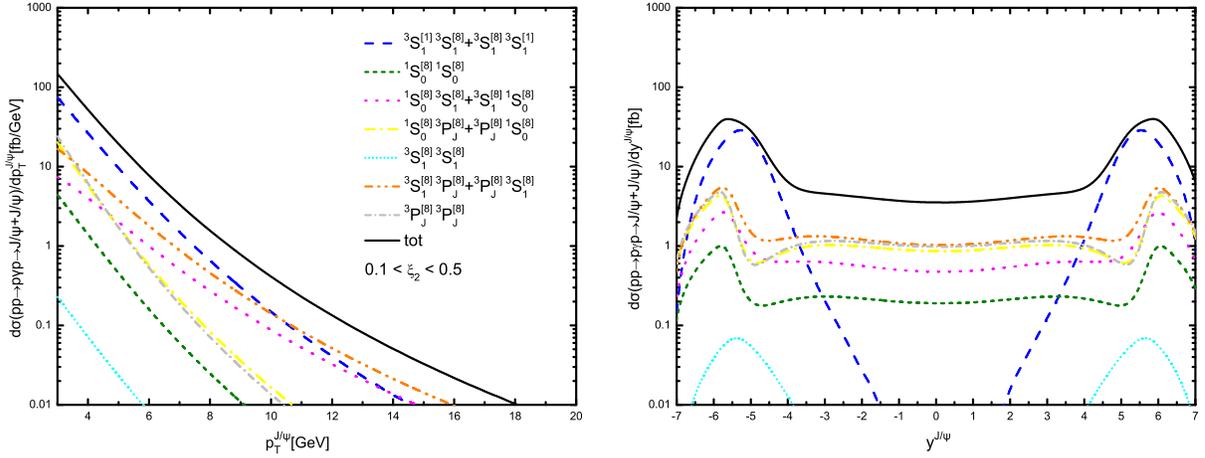}\newline
   \caption{ Same as in Fig. 4, but at the CMS-TOTEM forward detector with 0.1 $<$ $\xi_2$ $<$ 0.5. }
   \end{figure}

In Fig. 6, we give the LO distributions of $p_{T}^{J/\psi}$ (left panel) and $y^{J/\psi}$ (right panel) for the process pp $\rightarrow$ p$\gamma$p $\rightarrow$ $J/\psi$ + $J/\psi$ at the AFP-ATLAS forward detector with $\xi_3$. The behavior of the $p_{T}^{J/\psi}$ distribution is similar to that in Fig. 4; the $\sideset{^3}{^{[1]}_1}{\mathop{{S}}}$ + $\sideset{^3}{^{[8]}_1}{\mathop{{S}}}$ and $\sideset{^3}{^{[8]}_1}{\mathop{{S}}}$ + $\sideset{^3}{^{[1]}_1}{\mathop{{S}}}$ channels and $\sideset{^3}{^{[8]}_1}{\mathop{{S}}}$ + $\sideset{^3}{^{[8]}_J}{\mathop{{P}}}$ and $\sideset{^3}{^{[8]}_J}{\mathop{{P}}}$ + $\sideset{^3}{^{[8]}_1}{\mathop{{S}}}$ channels give the dominate contribution in the small and large $p_{T}^{J/\psi}$ region, respectively.  When 3 GeV $<$ $p_{T}^{J/\psi}$ $<$ 10 GeV, the d$\sigma/dp_{T}^{J/\psi}$ from the maximum value 1172.10 fb/GeV decreases to 2.51 fb/GeV. The total cross section can reach the magnitude of pb due to the enhancement by the effective photon density function in the small $\xi$ region.
\begin{figure}[!htbp]
   \includegraphics[scale=0.26]{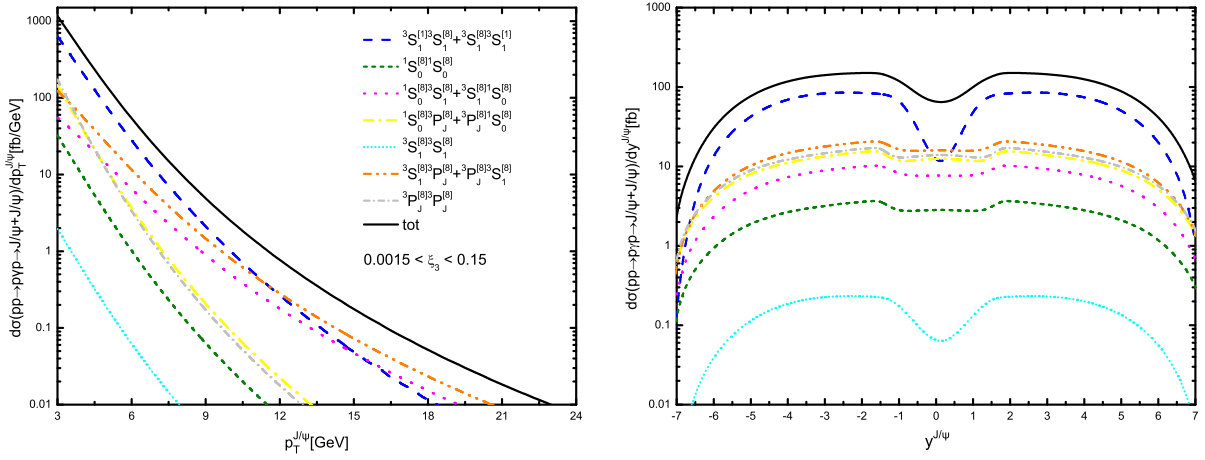}\newline
   \caption{ Same as in Fig. 4, but at the AFP-ATLAS forward detector with 0.0015 $<$ $\xi_3$ $<$ 0.15. }
   \end{figure}

To estimate the theoretical uncertainties caused by LDMEs, in Table II, we use three different sets of LDMEs to calculate the total cross section of double $J/\psi$ at the LHC with $\xi_1$, in which set 1 is extracted from the process of $J/\psi$\cite{A1} leptoproduction, and set 2 is extracted by the polarization of prompt $J/\psi$ at the Fermilab Tevatron\cite{T2}, set 3 is extracted though a joint fit to data on charmonium inclusive hadroproduction from runs I and II at the Fermilab Tevatron\cite{T1}. From this table, we can see that the total cross section of double $J/\psi$ slightly varies with different sets of LDMEs, and all the results are at the pb magnitude.

\begin{table}[!htbp]
\caption{The total cross section of $J/\psi$ with three sets of LDMEs, in which set 1 is extracted from the process of $J/\psi$\cite{A1} leptoproduction, set 2 is extracted by the polarization of prompt $J/\psi$ at the Fermilab Tevatron\cite{T2}, and set 3 is extracted though a joint fit to data on charmonium inclusive hadroproduction from runs I and II at the Fermilab Tevatron\cite{T1}. }
\begin{tabular}{|c|c|c|c|}
\hline\hline
        & Set 1 & Set 2  & Set 3  \\
\hline
  $<{\cal O}^{J/\psi}[^3S_1^{[1]}]>$ (GeV$^3$) & $1.1  $ & $1.4 $&          $1.4$ \\
  \hline
  $<{\cal O}^{J/\psi}[^1S_0^{[8]}]>$ (GeV$^3$) & $1 \times 10^{-2}$ & $3.3 \times 10^{-2}$& $3.65 \times 10^{-2}$ \\
  \hline
  $<{\cal O}^{J/\psi}[^3S_1^{[8]}]>$ (GeV$^3$)& $1.12 \times 10^{-2}$ & $3.9 \times 10^{-3}$ & $2.3 \times 10^{-3}$\\
  \hline
  $<{\cal O}^{J/\psi}[^3P_0^{[8]}]>$ (GeV$^5$)& $11.25 \times 10^{-3}$ &$2.18 \times 10^{-2}$& $2.25 \times 10^{-2}$ \\
  \hline
  $\sigma(pp \rightarrow p\gamma p \rightarrow J/\psi + J/\psi )$$\;$(fb)& 1369.06& 2493.94 &2527.90 \\
\hline\hline
\end{tabular}
\end{table}

We also calculated the cross section of double $\Upsilon$ for the process pp $\rightarrow$ p$\gamma$p $\rightarrow$ $\Upsilon$ + $\Upsilon$ at the LHC with forward detector acceptances. For three different forward detector acceptances, the total cross sections of double $\Upsilon$ at LO are 6.07, 0.83, and 5.57 fb with acceptances $\xi_1$, $\xi_2$, and $\xi_3$, respectively. The total cross section is too small and it is very difficult to detect the photoproduction of double $\Upsilon$ under the current experimental conditions. For illustration, the curves for the LO distributions of $p_{T}^{\Upsilon}$ and $y^{\Upsilon}$ for this process at the CMS-TOTEM forward detectors with $\xi_1$ are shown in Fig. 7.
\begin{figure}[!htbp]
   \includegraphics[scale=0.26]{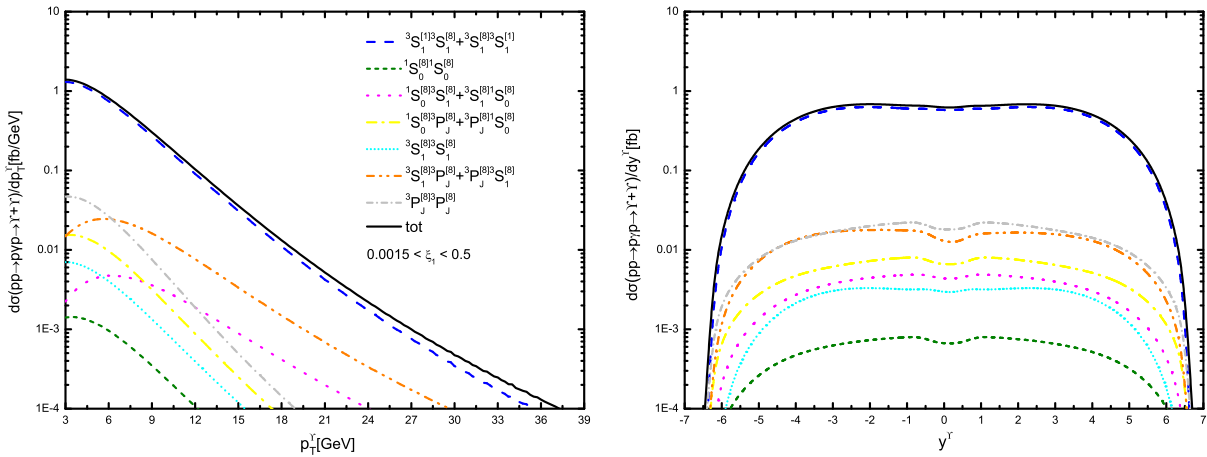}\newline
   \hspace*{-3cm}
   \caption{ Same as in Fig. 4, but for the process pp $\rightarrow$ p$\gamma$p $\rightarrow$ $\Upsilon$ + $\Upsilon$ at the CMS-TOTEM forward detector with 0.0015 $<$ $\xi_1$ $<$ 0.5.  }
   \end{figure}
\section{summary and discussion}
In this paper, we investigate the photoproduction of double $J/\psi$ ($\Upsilon$) to LO based on the NRQCD factorization formalism at the LHC with forward detector acceptances. We not only present the total cross section of double $J/\psi$ for pp $\rightarrow$ p$\gamma$p $\rightarrow$ $J/\psi$ + $J/\psi$ process at the different detector acceptances, but also give the numerical predictions of the differential cross section of the $p_{T}^{J/\psi}$ and $y^{J/\psi}$. For the process of double $J/\psi$ production, the $\sideset{^3}{^{[1]}_1}{\mathop{{S}}}$ + $\sideset{^3}{^{[8]}_1}{\mathop{{S}}}$ and $\sideset{^3}{^{[8]}_1}{\mathop{{S}}}$ + $\sideset{^3}{^{[1]}_1}{\mathop{{S}}}$ channels and $\sideset{^3}{^{[8]}_1}{\mathop{{S}}}$ + $\sideset{^3}{^{[8]}_J}{\mathop{{P}}}$ and $\sideset{^3}{^{[8]}_J}{\mathop{{P}}}$ + $\sideset{^3}{^{[8]}_1}{\mathop{{S}}}$ channels give the dominant contribution in the small and large $p_{T}^{J/\psi}$ region, respectively. The total cross section of double $J/\psi$ for the process pp $\rightarrow$ p$\gamma$p $\rightarrow$ $J/\psi$ + $J/\psi$ is less than 200 fb with $\xi_2$ since the effective photon density function is heavily suppressed at $\xi$ $>$ 0.1. When $\xi$ = $\xi_1$ and $\xi$ = $\xi_3$, the total cross section of double $J/\psi$ can reach the magnitude of pb since the effective photon density function is enhanced in the small $\xi$ region.  To estimate the cross section for the process pp $\rightarrow$ p$\gamma$p $\rightarrow$ $J/\psi$ + $J/\psi$ with the pure leptonic decays, we multiply the cross section for the direct production by the branching fraction at 12$\%$ for $J/\psi$ $\rightarrow$ $l^+l^-$. With the integrated luminosity of 200 fb$^{-1}$ at the LHC\cite{E5}, when we use the set 1 LDME, we could obtain about 3900, 500, and 3600 events for detector acceptances with $\xi_1$, $\xi_2$, and $\xi_3$, respectively. From our LO NRQCD predictions, we can see that the double $J/\psi$ photoproduction has the potential to be detected and may provide an interesting photoproduction signature at the LHC with forward detector acceptances, which will be useful for studying the mechanism of heavy quarkonium production. As we know, in the process of the heavy quarkonium production, especially in the process of double heavy charmonium production, the relativistic correction\cite{H1,H2,H3} and NLO QCD correction\cite{H4,H5,H6} may give substantial contributions; further investigations for relativistic correction and NLO QCD correction to processes of double charmonium photoproduction are meaningful works.  We also calculated the cross section of double $\Upsilon$ photoproduction at the LHC with different forward detector acceptances, but the total cross section is too small and it is very difficult to detect under the current experimental conditions.

\section*{Acknowledgments}
This work was supported in part by the National Natural
Science Foundation of China (Grants No. 11805001, No. 11305001, No. 11575002, No. 11675033, and No. 11747317)
and the Key Research Foundation of the Education Ministry of Anhui Province of China (Grant No. KJ2017A032).

\end{document}